# Simulations of Ion Velocity Distribution Functions Taking into Account Both Elastic and Charge Exchange Collisions

Huihui Wang[1,2], Vladimir S. Sukhomlinov[3], Igor D. Kaganovich[2], Alexander S. Mustafaev[4,5]

[1]School of Physical Electronics, University of Electronic Science and Technology of China, Chengdu, 610054, China

[2]Princeton Plasma Physics Laboratory, Princeton, NJ 08543, USA

[3]Department of Physics, St. Petersburg State University, St. Petersburg, 198504, Russia

[4]Department of General and Technical Physics, National Mineral-Resource University, St. Petersburg, 199106, Russia

[5]ITMO University, St. Petersburg, 197101, Russia

E-mail: huihuiwang@uestc.edu.cn; ikaganov@pppl.gov

**Abstract**

Based on accurate representation of the $He^+$-He angular differential scattering cross sections consisting of both elastic and charge exchange collisions, we performed detailed numerical simulations of the ion velocity distribution functions (IVDF) by Monte Carlo collision method (MCC). The results of simulations are validated by comparison with the experimental data of the ion mobility and the transverse diffusion. The IVDF simulation study shows that due to significant effect of scattering in elastic collisions IVDF cannot be separated into product of two independent IVDFs in the transverse and parallel to the electric field directions.

# 1. Introduction

The ion velocity distribution function (IVDF) plays a key part in the prediction and control of plasma parameters, especially for plasma etching [1], dust plasmas [2-4], auroral ionosphere [5-6] and Hall effect thruster [7-9]. In the previous works, IVDF is often calculated taking only charge exchange collisions into account without accounting for any scattering, and furthermore making simplifying assumption of a constant collision frequency [10-11] or a constant cross section [11-13].

However, scattering in the ion-atom collisions can be significant [14-16]. Therefore, ion-atom angular differential scattering cross sections [17-20] have to be taken into account for accurate calculations of IVDF.

To this end, we review effects of scattering in ion-atom collisions on IVDF formation. When describing ion-atom collisions, the following issues are frequently discussed:

a) Is it possible to separate ion-atom collisional process into elastic and charge exchange collisions [20-21]?

b) What is the effect of elastic ion-atom scattering on the ion mobility [11, 22] and IVDF? What is magnitude of error in IVDF associated with assumption of an isotropic elastic angular differential scattering [23-32]?

Technically, it is not possible to separate elastic collisions and charge-exchange collisions for collisions of ions and atoms of identical elements [20-21]. In this paper, we consider the angular differential cross sections of both elastic and charge-exchange processes as a whole. In section 2, a numerical model of the angular differential cross section is proposed. Based on this numerical model, a Monte Carlo method for ion-atom scattering is developed in section 3. Using the Monte Carlo method, IVDF is simulated in section 4. Finally, conclusions are presented in section 5.

# 2. Numerical model of the ion-atom angular differential cross section

Accurate calculations of the ion-atom angular differential cross section require making use of the quantum mechanical approach [19-21, 33-34], which shows that the cross section is not a sum of only elastic scattering and charge exchange processes. However, a simple model (semiclassical description) of ion-atom angular differential cross section proposed by McDaniel et al. [35] can be used for most transport processes. In this approach, the angular differential scattering cross section can be written in the form Eq. (1) [36], based on the assumption of classical nuclei trajectories which are not affected by electron exchange.

$$\sigma_\theta(\varepsilon, \theta) = P_{res}(\rho)\left(\frac{\rho}{\sin\theta'}\frac{d\rho}{d\theta'}\right)_{\theta'=\pi-\theta} + [1-P_{res}(\rho)]\frac{\rho}{\sin\theta}\frac{d\rho}{d\theta}, \quad (1)$$

where $\rho$ is the impact parameter and $P_{res}$ is the probability for electron transition from atom to ion, $\theta$ is the scattering angle in the reference frame of center mass, $\varepsilon = 0.5 m_{ion} m_{atom}/(m_{ion} + m_{atom}) \cdot (v_{ion} - v_{atom})^2$ is the energy of relative motion in the reference frame of center-of-mass, and $v_{ion}, v_{atom}, m_{ion}, m_{atom}$ are velocities and masses of ion and atom, respectively. Eq.(1) separates elastic and charge-exchange contributions to cross section explicitly. This approximation may yield some errors due to quantum mechanical effects, especially if the relative energy is small. In order to reduce this error, we consider the elastic and charge-exchange cross sections together as an inseparable and fit the momentum transfer cross section and the viscosity cross section for the total elastic and charge exchange cross sections together. As shown in the following, the fitting result for angular differential cross section agrees well with experiment data, although not reproducing the quantum-mechanical interference effects.

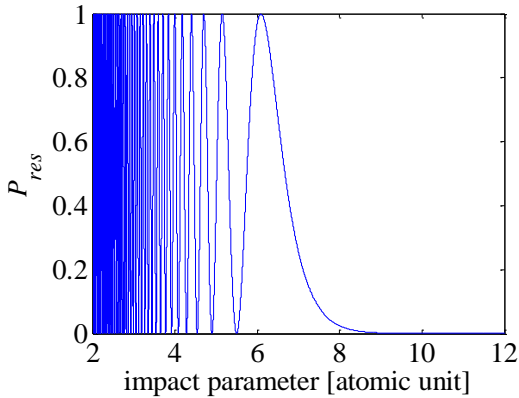

Figure 1. The charge-exchange probability of He$^+$+He at $\varepsilon$=1eV.

The function $P_{res}(\rho) = \sin^2\xi(\rho)$ is shown in Fig. 1 (in atomic units). The phase $\xi(\rho)$ is $v^{-1}[\pi/(2\gamma)]^{0.5}A^2\exp(-1/\gamma)\rho^{2/\gamma-1/2}\exp(-\rho\gamma)$, where $A$ and $\gamma$ are asymptotic parameters, and $v$ is the relative velocity in atomic unit [22]. For helium, $A$ is 2.87, $\gamma$ is 1.344. For small impact parameters, $P_{res}$ oscillates quickly between 0 and 1 with an average of 0.5 and decreases exponentially to 0 for large impact parameters (corresponding to small scattering angle $\theta_p$).

Function $\rho(\theta)$ can be determined from classical scattering of an ion on an atom with a polarization potential $U(r) \sim r^{-4}$. For such potential, the angular differential cross section is proportion to $1/(\theta_p^{1.5}\sin\theta_p)$ [15] for a small scattering angle. This function can be approximated as

$$\sigma_p(\varepsilon, \theta_p) = \frac{\rho}{\sin\theta}\frac{d\rho}{d\theta} \approx \frac{C}{[1-\cos\theta_p]^{1.25}}, \quad (2)$$

which is proportion to $1/\theta_p^{2.5}$ when $\theta_p \to 0$, with the same limit as $1/(\theta_p^{1.5}\sin\theta_p)$.

For simulations of ion transport in plasma, previous studies typically assumed a simplified model for ion-atom collisions: usually only assuming straight trajectory for charge exchange collisions [11], sometimes supplemented with isotropic elastic collisions [29-32]. For accurate simulations of IVDF we need to use a more accurate model of angular differential cross section for ion-atom collisions. Figure 2 shows experimental data for angular differential cross section for ion-atom collisions.

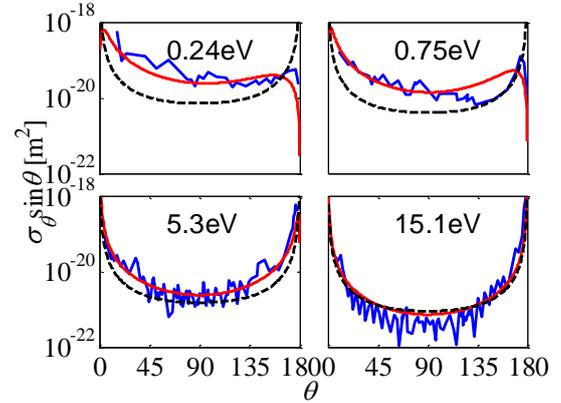

Figure 2. Angular differential cross sections. Experimental data are shown by the blue curve [18]. The red curve shows the approximation proposed in this paper, Eq.(3), and the black dashed is Phelps' model [37].

Combining Eq.(1) and Eq.(2), and assuming that at scattering angle $\theta \sim 1$ radian, the cross section is given by impact parameters where $P_{res} \approx 0.5$ as evident in Fig.1, we propose Eq.(3) to approximate the angular scattering differential cross section $\sigma_\theta(\varepsilon, \theta)$

$$\sigma_\theta(\varepsilon, \theta) = \frac{A(\varepsilon)}{[1-\cos\theta+a(\varepsilon)]^{1.25}} + \frac{A(\varepsilon)}{[1+\cos\theta+b(\varepsilon)]^{1.25}}, \quad (3)$$

where the first term describes the cross section for small-angle scattering, and the second term describes the cross section near $\pi$ angle. Small parameters, $a(\varepsilon)$ and $b(\varepsilon)$ are introduced to make the angular differential cross section integrable. We neglect interference terms, because as evident from experimental data two terms are sufficient for accurate description of the angular differential cross section. The parameters $A$, $a$, and $b$ in Eq.(3) are fitted to reproduce the total angular differential cross section (sum of elastic and charge-exchange cross sections), without its separation on elastic and charge-exchange collisions.

The total cross section, $\sigma_t$, the momentum transfer cross section, $\sigma_m$, and the viscosity cross section, $\sigma_v$, are calculated analytically making use of the approximation given by Eq.(3):

$$\sigma_t(\varepsilon) = 2\pi \int_0^\pi \sigma_\theta(\varepsilon, \theta) \sin\theta \, d\theta = 8\pi A \left[\frac{1}{a^{0.25}} - \frac{1}{(2+a)^{0.25}} + \frac{1}{b^{0.25}} - \frac{1}{(2+b)^{0.25}}\right], \quad (4)$$

$$\sigma_m(\varepsilon) = 2\pi \int_0^\pi \sigma_\theta(\varepsilon, \theta)(1-\cos\theta) \sin\theta \, d\theta = 8\pi A \left[\frac{a}{(2+a)^{0.25}} - \frac{4a^{0.75}}{3} + \frac{(2+a)^{0.75}}{3} - \frac{4(2+b)^{0.75}}{3} + \frac{2}{b^{0.25}} + \frac{4b^{0.75}}{3}\right], \quad (5)$$

$$\sigma_v(\varepsilon) = 2\pi \int_0^\pi \sigma_\theta(\varepsilon, \theta)(1-\cos^2\theta) \sin\theta \, d\theta =$$
$$8\pi A \left[\frac{2(2+a)^{0.75}}{3} - \frac{(2+a)^{1.75}}{7} - \frac{8a^{0.75}}{3} + \frac{5a(2+a)^{0.75}}{3} - \frac{32a^{1.75}}{21} + \frac{2(2+b)^{0.75}}{3} - \frac{(2+b)^{1.75}}{7} - \frac{8b^{0.75}}{3} + \frac{5b(2+b)^{0.75}}{3} - \frac{32b^{1.75}}{21}\right]. \quad (6)$$

The parameters $A$, $a$, and $b$ can be determined from the data for $\sigma_t$, $\sigma_m$ and $\sigma_v$ by solving Eqs. (4)-(6). For He$^+$+He cross sections, the approximations for $\sigma_t$, $\sigma_m$ and $\sigma_v$ have been developed for energies in the range between 0.01eV and 20eV according to data given in previous papers and are shown in Fig. 3.

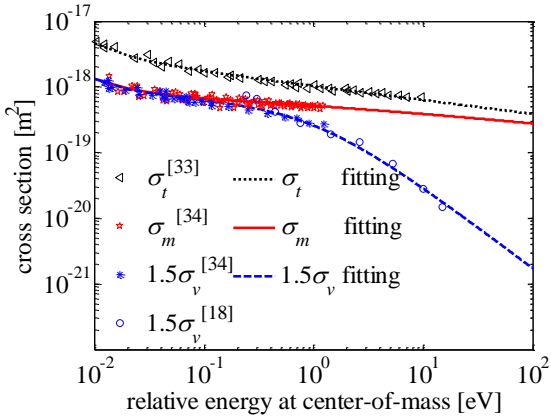

Figure 3. Approximate fit for the total, momentum, and viscosity cross sections for He$^+$+He collisions.

The fit for momentum cross section, $\sigma_m$, has been developed, for energy range (0.01eV~0.1eV) in Ref. [34] and for energy range ($\varepsilon$>0.1eV) in Ref. [38]

$$\sigma_m(\varepsilon) = 5.58 \times 10^{-19} \times [1 - 0.0557\ln(2\varepsilon)]^2 [1 + 0.0006\,\varepsilon^{-1.5}]. \quad (7)$$

Fit for total cross sections, $\sigma_t$, was developed making use of the theoretical calculation for the total cross section from Ref. [33]

$$\sigma_t(\varepsilon) = \sigma_m(\varepsilon)[1 + \varepsilon^{-0.2}]. \quad (8)$$

Fit for viscosity cross sections, $\sigma_v$, was obtained from theoretical calculation for the viscosity cross section [34] in the range (0.01~1eV) and the experimental data from Ref. [18] (1~20eV)

$$\sigma_v(\varepsilon) = \frac{\sigma_m(\varepsilon)}{1.5(1+\varepsilon^{1.1})}. \quad (9)$$

Given values of $\sigma_t$, $\sigma_m$ and $\sigma_v$, Eqs.(4)-(6) can be solved to obtain $A$, $a$, and $b$. If $a$ and $b$ are small relative to unity, then

$$A_0(\varepsilon) \approx \frac{21\sigma_v}{64\pi \times 2^{1.75}}, \quad (10)$$

$$b_0(\varepsilon) \approx \left(\frac{\sigma_m}{16\pi A_0} + \frac{1}{2^{0.25}}\right)^{-4}, \quad (11)$$

$$a_0(\varepsilon) \approx \left[\frac{\sigma_t}{8\pi A_0} + \frac{1}{2^{0.25}} + \frac{1}{(2+b_0)^{0.25}} - \frac{1}{b_0^{0.25}}\right]^{-4}. \quad (12)$$

where $A_0$, $a_0$, and $b_0$ are approximate values of $A$, $a$, and $b$.

However, in the range of $\varepsilon$ below 1 eV, $b$ is not very small relative to unity. The relative error of approximation given by Eqs.(10)-(12) is smaller than 1% at 4eV, while it reaches 17% at 0.01eV. Therefore, additional iterations can be performed to improve the accuracy of solution according to following iterative process:

$$A_{k+1}(\varepsilon) = \frac{\sigma_v}{8\pi}\left[\frac{2(2+a_k)^{0.75}}{3} - \frac{(2+a_k)^{1.75}}{7} - \frac{8a_k^{0.75}}{3} + \frac{5a_k(2+a_k)^{0.75}}{3} - \frac{32a_k^{1.75}}{21} + \frac{2(2+b_k)^{0.75}}{3} - \frac{(2+b_k)^{1.75}}{7} - \frac{8b_k^{0.75}}{3} + \frac{5b_k(2+b_k)^{0.75}}{3} - \frac{32b_k^{1.75}}{21}\right]^{-1}, \quad (13)$$

$$b_{k+1}(\varepsilon) = 16\left\{\frac{\sigma_m}{8\pi A_{k+1}} - \frac{a_k}{(2+a_k)^{0.25}} + \frac{4a_k^{0.75}}{3} - \frac{(2+a_k)^{0.75}}{3} + \frac{4(2+b_k)^{0.75}}{3} - \frac{4b_k^{0.75}}{3}\right\}^{-4}, \quad (14)$$

$$a_{k+1}(\varepsilon) = \left\{\frac{\sigma_t}{8\pi A_{k+1}} + \frac{1}{(2+a_k)^{0.25}} + \frac{1}{(2+b_{k+1})^{0.25}} - \frac{1}{b_{k+1}^{0.25}}\right\}^{-4}. \quad (15)$$

Eqs. (13)-(15) are used for $A$, $b$, $a$, respectively. Given coefficients $A_k$, $b_k$ and $a_k$, the cross sections can be calculated $\sigma_{t\_k}$, $\sigma_{m\_k}$, and $\sigma_{v\_k}$.

The maximum relative error $|(\sigma_{t\_k}-\sigma_t)/\sigma_t|$, $|(\sigma_{m\_k}-\sigma_m)/\sigma_m|$, and $|(\sigma_{v\_k}-\sigma_v)/\sigma_v|$ are presented in Table 1, which shows the relative error is smaller than $5.7\times10^{-4}$ after 7 iterations. Therefore, values

of $A_7$, $b_7$ and $a_7$ are adopted for approximation of cross sections in this paper, as shown in figure 4.

Given the functions of $A$, $a$, and $b$, determined from $\sigma_t$, $\sigma_m$ and $\sigma_v$, the angular differential cross sections of both scattering and charge exchange processes are calculated and compared to experimental data [18] and Phelps' model [37], as shown in Fig.2. Phelps' model assumes symmetry ($P_{res} = 0.5$) regarding transformation ($\theta \rightarrow \pi - \theta$) in Eq.(1), and makes use of dipole polarizabilities instead of $\sigma_t$, $\sigma_m$, and $\sigma_v$.

Table 1. The maximum relative error during [0.01eV, 20eV]

|  | $k=0$ | $k=1$ | $k=3$ | $k=5$ | $k=7$ |
|---|---|---|---|---|---|
| $\sigma_{t\_k}$ | $3.0 \times 10^{-5}$ | $2.3 \times 10^{-5}$ | $4.9 \times 10^{-6}$ | $1.0 \times 10^{-6}$ | $2.1 \times 10^{-7}$ |
| $\sigma_{m\_k}$ | $2.4 \times 10^{-2}$ | $1.1 \times 10^{-2}$ | $2.4 \times 10^{-3}$ | $5.4 \times 10^{-4}$ | $1.2 \times 10^{-4}$ |
| $\sigma_{v\_k}$ | $1.7 \times 10^{-1}$ | $6.1 \times 10^{-2}$ | $1.2 \times 10^{-2}$ | $2.5 \times 10^{-3}$ | $5.7 \times 10^{-4}$ |

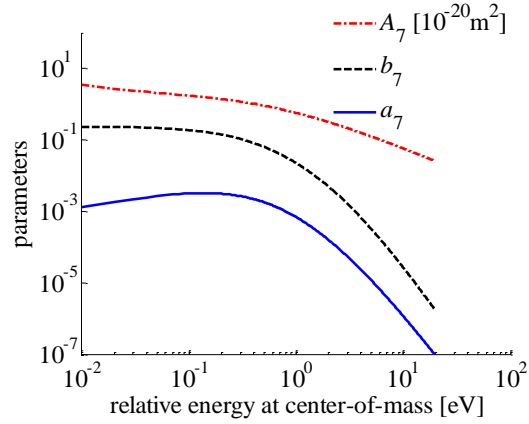

Figure 4. $A_7$, $b_7$ and $a_7$ parameters as functions of energy of relative motion of ion and atom in the center mass reference frame.

## 3. Monte Carlo collision model of ion-atom scattering

For convenience of implementation, we separate the collision process into two processes according to Eq.(3). The total cross sections of the first process and the second process in the right-hand side of Eq.(3) are

$$\sigma_{t\_1}(\varepsilon) = 2\pi \int_0^\pi \sigma_1(\varepsilon, \theta) \sin\theta \, d\theta = \frac{8\pi A}{a^{0.25}} - \frac{8\pi A}{(2+a)^{0.25}}, \quad (16)$$

$$\sigma_{t\_2}(\varepsilon) = 2\pi \int_0^\pi \sigma_2(\varepsilon, \theta) \sin\theta \, d\theta = \frac{8\pi A}{b^{0.25}} - \frac{8\pi A}{(2+b)^{0.25}}, \quad (17)$$

where $\sigma_1(\varepsilon, \theta)$ and $\sigma_2(\varepsilon, \theta)$ are the first item and the second item in the right-hand side of Eq.(3), respectively.

After collision, due to conservation of momentum and energy, the ion velocities are changed according to Eqs.(18)-(19) [39-40].

$$\begin{cases} \boldsymbol{v}_{\alpha\_after} = \boldsymbol{v}_\alpha + \frac{m_{\alpha\beta}}{m_\alpha} \Delta\boldsymbol{u} \\ \boldsymbol{v}_{\beta\_after} = \boldsymbol{v}_\beta - \frac{m_{\alpha\beta}}{m_\beta} \Delta\boldsymbol{u} \end{cases} \quad (18)$$

$$\begin{cases} \Delta u_x = \left(\frac{u_x}{u_\perp}\right) u_z \sin\theta \cos\Phi - \left(\frac{u_y}{u_\perp}\right) u \sin\theta \sin\Phi - u_x(1-\cos\theta) \\ \Delta u_y = \left(\frac{u_y}{u_\perp}\right) u_z \sin\theta \cos\Phi + \left(\frac{u_x}{u_\perp}\right) u \sin\theta \sin\Phi - u_y(1-\cos\theta) \\ \Delta u_z = -u_\perp \sin\theta \cos\Phi - u_z(1-\cos\theta) \end{cases}$$

(19)

where $m_{\alpha\beta}$ is $m_\alpha m_\beta/(m_\alpha + m_\beta)$, $\boldsymbol{u}$ is $\boldsymbol{v}_\alpha - \boldsymbol{v}_\beta$, $u_\perp$ is $(u_x^2 + u_y^2)^{0.5}$, $\boldsymbol{v}_\alpha$ and $\boldsymbol{v}_\beta$ are the velocities before collision, $\boldsymbol{v}_{\alpha\_after}$ and $\boldsymbol{v}_{\beta\_after}$ are the velocities after collision, $\Phi$ is an azimuthal scattering angle [0, $2\pi$], and $\theta$ is the polar scattering angle [0, $\pi$]. The value of $\theta$ is according to the cumulative probability distribution derived from the angular differential cross section [29, 41-42].

Introducing $R_1$ and $R_2$, uniform random numbers between 0 and 1 for both processes,

$$R_{1,2}(\theta) = \frac{\int_0^{\theta_{1,2}} \sigma_{1,2}(\varepsilon, \theta) \sin\theta \, d\theta}{\int_0^\pi \sigma_{1,2}(\varepsilon, \theta) \sin\theta \, d\theta}, \quad (20)$$

according to the cumulative probability distribution Eq.(20), the polar scattering angles for MCC are obtained

$$\cos\theta_1 = 1 + a - \{a^{-0.25} - R_1[a^{-0.25} - (2+a)^{-0.25}]\}^{-4}, \quad (21)$$

$$\cos\theta_2 = -(1+b) + \{(2+b)^{-0.25} + R_2[b^{-0.25} - (2+b)^{-0.25}]\}^{-4}. \quad (22)$$

Table 2. Three models for angular differential cross sections

| | $\sigma_\theta(\varepsilon, \theta)$ | $\sigma_m$ | $\sigma_v$ |
|---|---|---|---|
| Model 1 | $\dfrac{A}{(1 - \cos\theta + a)^{1.25}} + \dfrac{A}{(1 + \cos\theta + b)^{1.25}}$ | Eq.(7) | Eq.(9) |
| Model 2 | $\dfrac{\sigma_m}{2} \dfrac{\delta(\theta - \pi)}{2\pi \sin\theta}$ | Eq.(7) | 0 |
| Model 3 | $\dfrac{\sigma_i}{4\pi} + \sigma_b \dfrac{\delta(\theta - \pi)}{2\pi \sin\theta}$ | Eq.(7) | Eq.(9) |

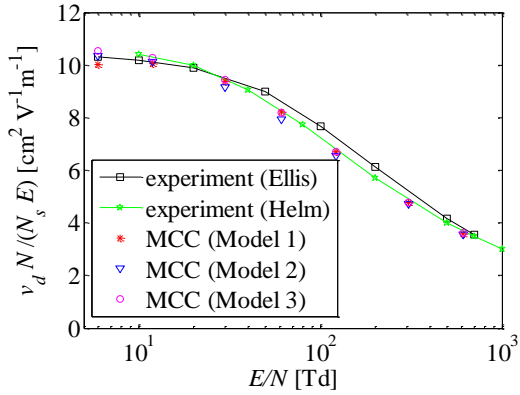
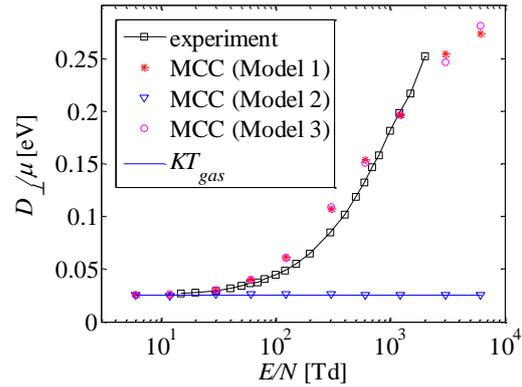

Figure 5. The product of mobility and gas pressure for He$^+$+He.

Figure 6. The relation between $D_\perp/\mu$ and $E/N$ for He$^+$+He.

## 4. The ion velocity distribution functions simulated by Monte Carlo Collision method

Based on the approximation for the angular differential cross section developed in Sec.2 (Model 1), IVDFs are simulated for helium discharges at 0.1Torr pressure and 294K gas temperature. We compared the results for IVDFs with predictions of less accurate models, where only charge-exchange collisions were taken into account without taking into account any scattering process (Model 2). In Model 3 isotropic elastic scattering was added to Model 2 as specified in Table 2.

The functions $(A, a, b, \sigma_i, \sigma_b)$ used in models for angular differential cross section are expressed through the integrated cross sections $\sigma_m, \sigma_v, \sigma_t$, which are known from experimental data or quantum-mechanical calculations. Model 1 is the proposed model in this paper described by Eq.(3); the angular differential cross section given by Eq.(3) agrees well with the data obtained in experiments with ion beams. Model 2 is the widely-used model, see e.g. Refs. [10-13], which takes into account only the resonant charge-exchange collisions (without the scattering of the ion in the polarization potential). Thus, according to Model 2, ions and atoms move along straight lines during a collision and the scattering angle in the center mass reference frame is exactly π. Finally, Model 3 considers two processes: the isotropic elastic scattering in the center mass reference frame with $\sigma_i \equiv 1.5\sigma_v$ and the backward collisions with the angular differential scattering cross section at an angle of π in proportion to the delta-Dirac's function with $\sigma_b \equiv (\sigma_m - \sigma_i)/2$, see e.g. Refs. [30-32].

The ion transport properties predicted by these three models are examined by comparing with the experimental data from Refs. [43-45] shown in Figs. 5-6, where $v_d$ is the ion drift velocity, $E$ is the electric field, $N$ is the gas density, $N_s$=2.6868×10$^{19}$ cm$^{-3}$ is the standard gas number density, $\mu$ is the mobility ($v_d/E$), and $D_\perp$ is the transverse diffusion coefficient. In simulations, $D_\perp$ is measured using relation $<r_\perp^2>$=4$D_\perp t$, where $<r_\perp^2>$ is the mean square of the transverse distance from the origin [30]. Figure 5 shows that the mobility predicted by all three models are in agreement with the experimental data, because the momentum transfer cross section responsible for the drift velocities is accurately described in all these models. In addition, figure 5 also shows that our simulation results agree well with the Helm's experimental data [44] and do not agree with the Ellis' data [43]. This is consistent with the conclusion of Ref. [46], in which the authors claim that the Helm's data is more accurate than previous experimental data. Figure 6 shows that the transverse diffusion coefficient calculated based on Model 1 and Model 3 is consistent with the experimental data [45], because scattering process and energy transfer between the transverse and parallel directions as described by the viscosity cross section is adequately described in these models, whereas Model 2 gives inaccurate transverse diffusion coefficient, because it completely neglects scattering process and energy transfer between the transverse and parallel directions (and the viscosity cross section is exactly zero in this Model). Therefore, the value of $D_\perp/\mu = KT_{gas}$ is independent of the electric field in Model 2.

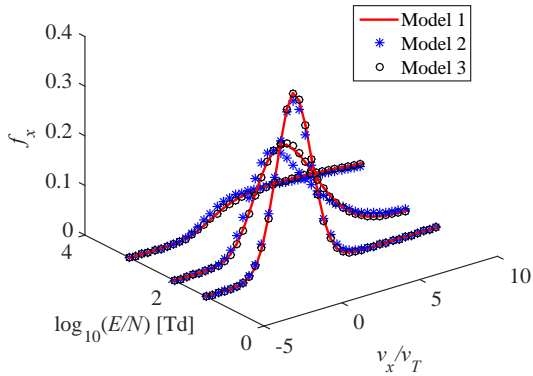

Figure 7. IVDF of He$^+$ ions in the parallel direction.

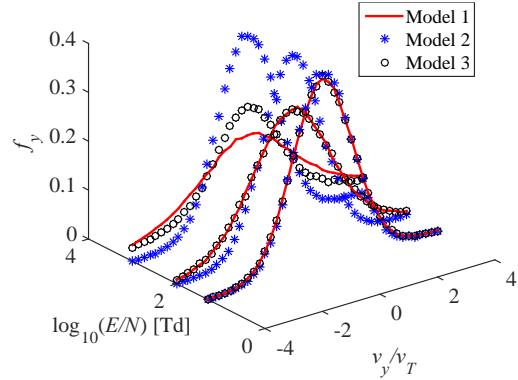

Figure 8. IVDF He$^+$ ions in the transverse direction.

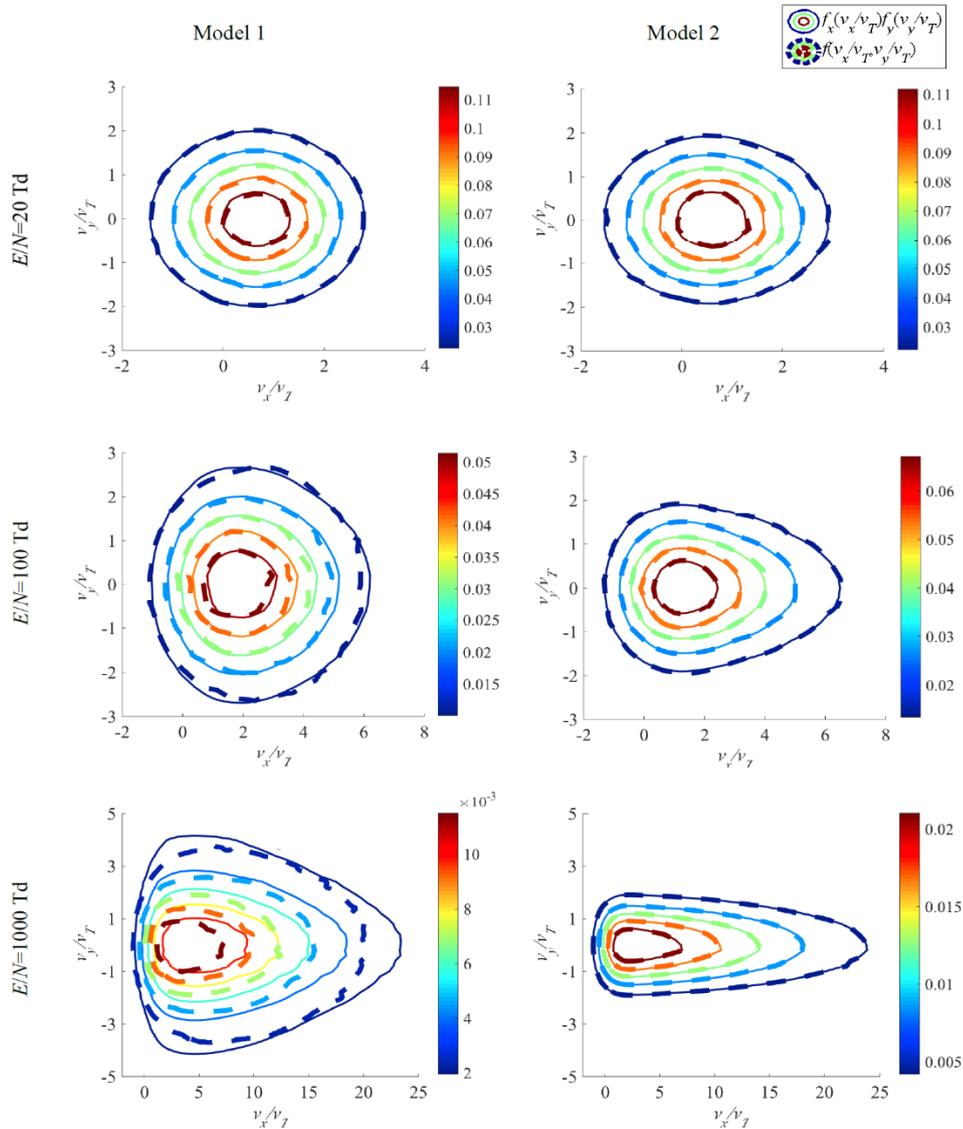

Figure 9. IVDF of He$^+$ ions in parallel and perpendicular directions for different values of $E/N$.

IVDFs in different directions relative to the electric field are simulated for different values of reduced electric field $E/N$=20Td, 100Td and 1000Td. $f_x$ is the IVDF for velocity direction parallel to the electric field, $f_y$ is the IVDF for velocity direction perpendicular to the electric field, and $v_T$ is the atom thermal velocity for gas temperature of 294K.

IVDFs of He$^+$ ions in the direction parallel to the electric field, $f_x$, are almost identical in all three models as shown in figure 7 (because of making use of the same accurate momentum transfer cross section). IVDFs of He$^+$ ions in the transverse direction, $f_y$, are different in all three models as shown in Fig. 8. Model 2 predicts $f_y$ as the Maxwellian distribution with the ion temperature equal to the gas temperature, $T_{gas}$, because of absence of scattering. Whereas $f_y$ in Model 1 and Model 3 are gradually deviating from the Maxwellian distribution with $T_{gas}$ for higher values of $E/N$, because of the energy transfer between different directions. Note that if $E/N$ becomes higher than 1000Td, IVDFs, $f_y$ in Model 1 and Model 3 begin to deviate from each other significantly.

Because of the symmetry between $y$ and $z$ directions, IVDF can be represented by the two-dimensional velocity distribution function $f(v_x, v_y)$. If IVDFs are independent in different directions, then $f(v_x, v_y) = f(v_x)f(v_y)$. However, there may be correlations between different directions, which makes IVDF much more complex function than $f(v_x)f(v_y)$.

Figure 9 shows the two-dimensional velocity distributions $f(v_x, v_y)$ for three values of $E/N$. Similarly to results obtained in Ref. [11], $f(v_x, v_y)$ can be separated into the product of $f_x$ and $f_y$ only for Model 2. The property of IVDF $f(v_x, v_y)$ that it can be represented as a product of two independent IVDFs $f_x(v_x) \times f_y(v_y)$ is based on that ion velocity directions stay the same after collisions in Model 2 (scattering angle $\pi$ in the reference frame of the center of mass). However, accounting for angular scattering breaks this property for Model 1 and Model 3. (This phenomenon is similar in Model 1 and Model 3, therefore we only present the results of Model 1 in figure 9). The anisotropy of IVDF increases with $E/N$.

Figure 8 show that the difference in IVDF for Model 1 and 3 emerges only for sufficiently strong electric fields. Therefore, we show details of IVDF calculated with different models for high $E/N$=1000Td in Fig.10. Figure 10(a) shows IVDF of Model 2 is consistent with the previously obtained theoretical result of Ref. [13]. Figure 10(b) shows the difference between IVDFs obtained using Model 1 and Model 3, which are both more isotropic than IVDF given by Model 2.

The difference between IVDFs obtained with Model 1 and Model 3 is caused by their different differential cross section, which is demonstrated in figure 11. The cross section near 90 degree in Model 3 is larger than that in Model 1, which means ions after collision in Model 3 has a bigger probability to acquire a large transverse energy, $w_y$. This property makes $f_y$ at a large ion transverse speed in Model 3 is slightly larger than that in Model 1. Because of the approximately equal average transverse ion energy due to the same transport cross sections $\sigma_m$ and $\sigma_v$, there should be at least two intersection points between the $f_y$ curves of Model 1 and Model 3. This phenomenon is shown in figure 12, where $w_y$ is defined as $m_{ion}v_y^3/(2|v_y|)$. Figure 12 also clearly shows $f_y$ in Model 1 deviates from Maxwellian distribution (i.e. straight line), which is consistent with the conclusion of Ref. [46].

The normalized ion energy distribution functions (IEDF) for $E/N$=1000Td obtained with different models are shown in figure 13 (a). The ion energy for the peak of IEDF is in the range of 0.02~0.04eV, which is of the order of the atom temperature $KT_{gas}$=0.025eV. The difference in the IEDFs can be explained by the different ion energies obtained after collisions in different models: in Model 2, energies of all ions after collisions are determined by energies of atoms due to the charge exchange collisions; in Model 3, ion energies of a part of ions after collisions are reduced to energies of atoms due to the backward collisions; in Model 1, on the contrary, the scattering of ions and atoms in collisions with the polarization potential yields relatively high ion energies compared to atom energies. Because of this, IEDF at the ion energy with the order of the atom temperature $KT_{gas}$=0.025eV is highest in Model 2, and lowest in Model 1. Besides, IEDFs from different models have at least two intersection points because of the approximately equal average ion energy due to the same cross sections $\sigma_m$, as shown in figure 13 (b), where IEDF$_1$, IEDF$_2$, IEDF$_3$ are IEDF in Model 1, Model 2, Model 3, respectively.

Figures 14 and 15 show the details of the angular distribution functions. The average angle $\theta_{average}$ is defined as Eq.(23),

$$\theta_{average}(\varepsilon_{ion}) = \frac{\int_0^\pi \theta F(\varepsilon_{ion},\theta) \sin\theta \, d\theta}{\int_0^\pi F(\varepsilon_{ion},\theta) \sin\theta \, d\theta}, \quad (23)$$

where $\theta$ is the angle between the ion velocity and the electric field, and $F(\varepsilon_{ion},\theta)$ is the energy and angle distribution function normalized by Eq.(24).

$$\int_0^{+\infty}\int_0^{\pi} F(\varepsilon_{ion},\theta)\sin\theta\, d\theta\, d\varepsilon = 1 \qquad (24)$$

Larger $\theta_{average}$ means IVDF is more isotropic. Figures 14 shows that $\theta_{average}$ decreases with increase of ion energy for all models. $\theta_{average}$ in Model 1 has the highest value, $\theta_{average}$ in Model 2 has the lowest value, and $\theta_{average}$ in Model 3 lies in between Model 2 to Model 1 predictions. One example of the angular distribution for $\varepsilon_{ion}$=0.1eV is shown in figure 15, which also shows that the angular distribution given by Model 1 is mostly isotropic and the angular distribution is most anisotropic for Model 2.

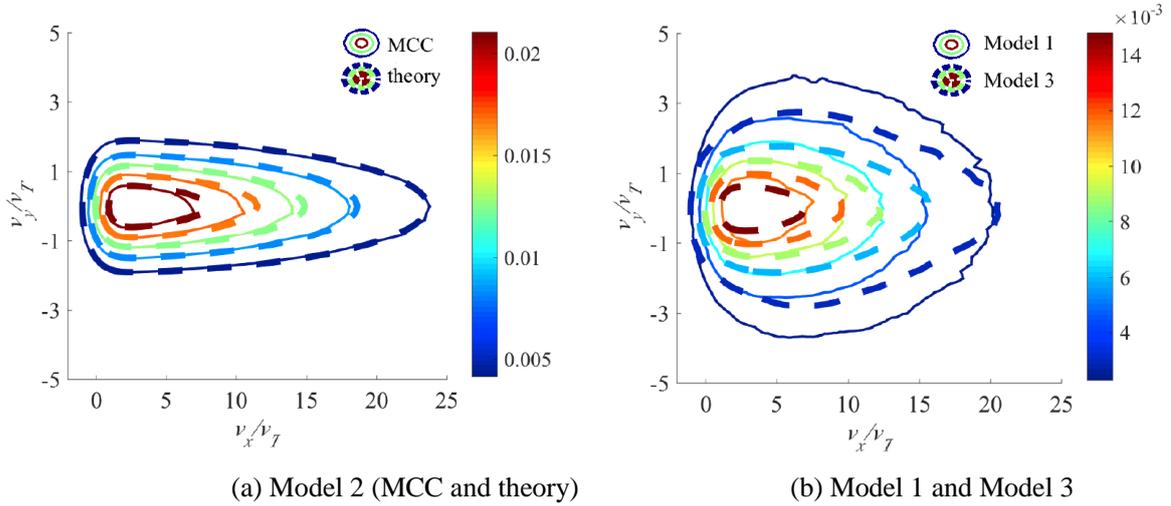

(a) Model 2 (MCC and theory)  (b) Model 1 and Model 3

Figure 10. The contour plot of IVDF $f(v_x/v_T, v_y/v_T)$ of He$^+$ ions at $E/N$=1000Td.

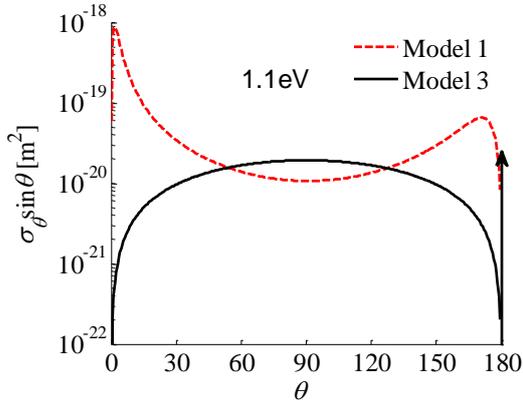 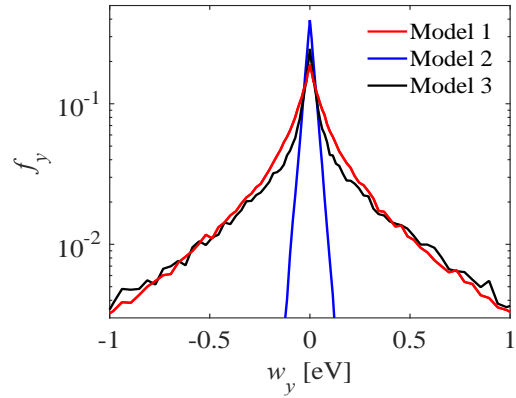

Figure 11. Angular differential cross sections of various models.   Figure 12. $f_y$ at $E/N$=1000Td.

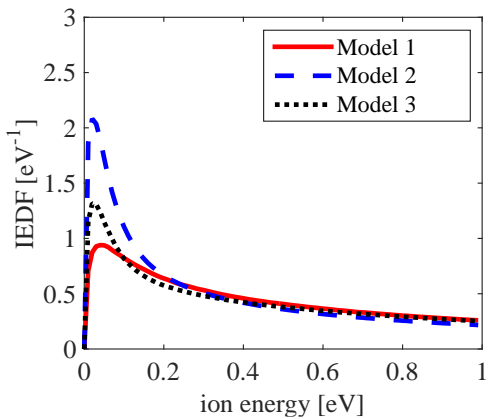 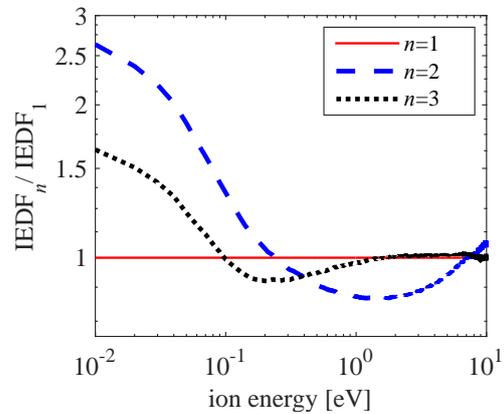

(a) IEDF  (b) the ratio of IEDFs

Figure 13. Ion energy distribution functions at $E/N$=1000Td.

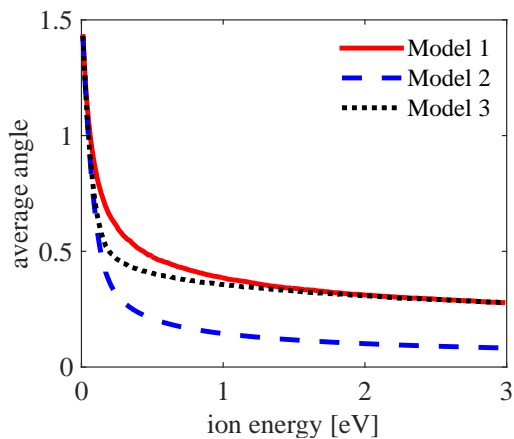

Figure 14. Average angle at $E/N$=1000Td.

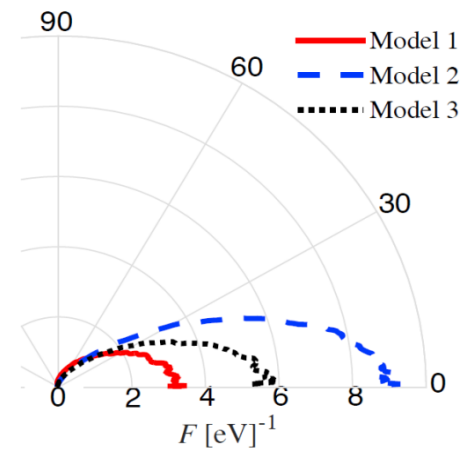

Figure 15. Angular distribution at $\varepsilon_{ion}$=0.1eV.

## 5. Conclusions

Based on the developed fit for ion-atom angular differential scattering cross sections, MCC model is proposed for simulation of IVDF in helium discharges. The predictions of the model are compared to other models used in the literatures. We show that taking into consideration both elastic and charge exchange collisions rather than ignoring the elastic collisions is important for correct simulation of IVDF, when there is a requirement of high-precision calculation of IVDF in the transverse to the electric field direction. The fit method for ion-atom angular differential scattering cross sections developed in this paper makes use of the total, momentum, and viscosity cross sections can be applied to other gases. Based on the developed model, the follow-up paper (Ref. [47]) compares IVDF obtained in simulations with recent experimental data of Refs. [13, 48].

## Acknowledgment:

We are thankful to Alexander V. Khrabrov for fruitful discussions and benchmarking of the Monte Carlo code used in the paper with EDIPIC code and Predrag Krstic for valuable discussions on cross sections. The work of H. Wang was supported by China Scholarship Council and the Fundamental Research Funds for the Central Universities ZYGX2014J040, and Igor D. Kaganovich was supported by the U.S. Department of Energy.